 \documentstyle[11pt]{article}
 
\newcommand{\be}{\begin{equation}}
\newcommand{\ee}{\end{equation}}
\newcommand{\bea}{\begin{eqnarray}}
\newcommand{\eea}{\end{eqnarray}}
\newcommand{\beann}{\begin{eqnarray*}}
\newcommand{\eeann}{\end{eqnarray*}}
\newcommand{\beasn}{\begin{sneqnarray}}
\newcommand{\eeasn}{\end{sneqnarray}}
\newcommand{\bref}[1]{(\ref{#1})}

\newcommand{\NPB}[3]{{\sl Nucl. Phys.} {\bf B#1} (19#2)  {#3}}

\newcommand{\PLB}[3]{{\sl Phys. Lett.} {\bf #1B} (19#2)  {#3}}

\newcommand{\IJA}[3]{{\sl Int. J. Mod. Phys.} {\bf A#1} (19#2) {#3}}
\newcommand{\CMP}[3]{{\sl Commun. Math. Phys.} {\bf #1} (19#2) {#3}}

\catcode`@=11
\def\dif{{\rm d}}
\def\deriv{\@ifnextchar[{\@deriv}{\@deriv[]}}
   \def\@deriv[#1]#2#3{\mathchoice%
{{\dif^{#1}#2\over\dif{#3}^{#1}}}{{\dif^{#1}#2/\dif{#3}^{#1}}}%
{{\dif^{#1}#2\over\dif{#3}^{#1}}}{{\dif^{#1}#2/\dif{#3}^{#1}}}}

\def\presup#1{{}^{#1}\kern-.15em\relax}      %pre-superscript
\def\presub#1{{}_{#1}\kern-.12em\relax}      %pre-subscript

\def\secteqno{\@addtoreset{equation}{section}%
\def\theequation{\thesection.\arabic{equation}}}
\def\endsecteqno{\def\theequation{\@ifundefined{chapter}%
{\arabic{equation}}{\thechapter.\arabic{equation}}}}
\newcounter{subequation}
\def\thesubequation{\alph{subequation}}
\def\sneqnarray{\stepcounter{equation}\let\@currentlabel=\theequation
\setcounter{subequation}{1}
\def\@eqnnum{{\rm (\theequation\thesubequation)}}
\global\@eqcnt\z@\tabskip\@centering\let\\=\@eqncr\let\@@eqncr=\@@sneqncr
$$\halign to \displaywidth\bgroup\@eqnsel\hskip\@centering
 $\displaystyle\tabskip\z@{##}$&\global\@eqcnt\@ne
 \hskip 2\arraycolsep \hfil${##}$\hfil
 &\global\@eqcnt\tw@ \hskip 2\arraycolsep $\displaystyle\tabskip\z@{##}$\hfil
  \tabskip\@centering&\llap{##}\tabskip\z@\cr}
\def\endsneqnarray{\@@sneqncr\egroup $$\global\@ignoretrue}
\def\@@sneqncr{\let\@tempa\relax
   \ifcase\@eqcnt \def\@tempa{& & &}\or \def\@tempa{& &}
   \else \def\@tempa{&}\fi
     \@tempa \if@eqnsw\@eqnnum\stepcounter{subequation}\fi
     \global\@eqnswtrue\global\@eqcnt\z@\cr}
\def\nobiblabels{\def\@lbibitem[##1]##2{\@bibitem{##2}}}
\catcode`@=12

\def\pt{\partial_t}
\def\pz{\partial_z}

\def\p{\partial}

\def\gttt{\Gamma_{tt}^t}
\def\gttz{\Gamma_{tz}^t}

\def\gtzz{\Gamma_{zz}^t}

\def\gztt{\Gamma_{tt}^z}
\def\gztz{\Gamma_{tz}^z}

\def\gzzz{\Gamma_{zz}^z}

\def\et{\epsilon^t}
\def\ez{\epsilon^z}

\def\ja{\left| {\bf J} \right|}

\def\W{$\cal W$}

\secteqno
\title{{\bf Finite {\W}$_3$ Transformations in a Multi-time Approach}}
\author{{\sc J. Gomis}$^\spadesuit$,
        {\sc J. Herrero}$^\spadesuit$,
        {\sc K. Kamimura}$^\clubsuit$
        {\sc and J. Roca}$^\diamondsuit$\\
        \llap{$^\spadesuit$}%
        \small{\it{Departament d'Estructura i Constituents
               de la Mat\`eria}}\\
        \small{\it{Universitat de Barcelona and }}\\
        \small{\it{Institut de F\'{\i}sica d'Altes Energies}}\\
        \small{\it{Diagonal, 647}}\\
        \small{\it{E-08028 BARCELONA}}\\
        \llap{$^\clubsuit$}%
        \small{\it{Department of Physics, Toho University}}\\
        \small{\it{Funabashi}}\\
        \small{\it{274 JAPAN}}\\
        \llap{$^\diamondsuit$}%
        \small{\it{Department of Physics, Queen Mary and Westfield
College}}\\
        \small{\it{Mile End Road, London E1 4NS}}\\
        \small{\it{ENGLAND}}\\
        {\it e-mails:} \small{GOMIS@EBUBECM1, HERRERO@EBUBECM1,}\\
                       \small{KAMIMURA@JPNYITP, J.ROCA@QMW.AC.UK}}

\date{}

\begin{document}

\maketitle

\thispagestyle{empty}

\begin{abstract}
Classical {\W}$_3$ transformations are discussed as restricted
diffeomorphism transformations (\W-Diff) in two-dimensional space.
We formulate them by using Riemannian geometry as a basic ingredient.
The extended {\W}$_3$ generators are given as particular combinations of
Christoffel symbols. The defining equations of \W-Diff are shown to
depend on these generators explicitly.  We also consider the issues of
finite transformations, global $SL(3)$ transformations and \W-Schwarzians.
\end{abstract}

\vskip 15mm

\vfill
\vbox{
\hfill August 1994\null\par
\hfill UB-ECM-PF 94/20\null\par
\hfill TOHO-FP-9448\null\par
\hfill QMW-PH-94-25}\null

\clearpage

%%%%%%%%%%%%%%%%%%%%%%%%%%%%%%%%%%%%%%%%%%%%%%%%%%%%%%%%%%%%%%%
%\baselineskip 15pt

\section{Introduction}

\hspace{\parindent}%
Non-linear extensions of the Virasoro algebra are known as
\W-algebras \cite{Z}.
The classical counterparts of these algebras are obtained through
Drinfel'd-Sokolov (DS) Hamiltonian reduction for Kac-Moody current
algebras \cite{DS} and the zero-curvature approaches
\cite{P}. Classical \W-algebras are also related to
the theory of integrable systems through the Gel'fand-Dickey brackets
\cite{Dickey}. For example,
classical {\W}$_N$ transformations are the symmetries of the equation
\be
\label{DSeq}
LX\equiv
\partial^N X\,+\,\sum_{j=0}^{N-2}U_{N-j}\,\,\partial^j\,X\,=\,0,
\ee
 where $X$ and $U_j$  fields  are regarded as functions of a single
variable $t$ \cite{Radul}. The reparametrizations of $t$ induce
diffeomorphism transformations (Diff) on $X$ and $U_j$, whereas
\W-transformations cannot be understood as reparametrizations in
one-dimensional base space. In order to understand all
\W-transformations as diffeomorphisms it is natural to extend the
base space to a multi-dimensional one (\W-space) \cite{Matsuo}.

In \W-space we will consider multi-time extension equations
associated with \bref{DSeq} given by:
\bea
\nonumber
& LX = 0,
\\
\label{multiDSeq}
& \partial_{t_k}X=L^{k/N}_+ X, \quad\quad\quad k=2,\ldots,N-1,
\eea
where $X$ and $U_j$ depend now on $N-1$  time parameters:\,
$t^1,\ldots,t^{N-1}$ and $L^{k/N}_+$
 is the differential part of the pseudo-differential
operator
$L^{k/N}$ \cite{Dickey}. Therefore $X$ is a truncation of the
Baker-Akhiezer
function associated to the generalized hierarchies of partial
differential equations generated by $L$.
In this framework the {\W}$_N$ transformations are those leaving
the whole set of multi-time equations \bref{multiDSeq} invariant.
They are induced by some restricted class of Diff in \W-space which we
refer to as \W-Diff.

We will also show how Riemannian
geometry is useful for the formulation of the multi-time equations.
We derive them from a set of covariant equations by imposing
suitable gauge-fixing conditions on  the Christoffel symbols.
In particular for $N=3$ we will explicitly see how {\W}$_3$
generators appear as some combinations of the Christoffel symbols
and how they transform under finite \W-Diff.
The {\W}-Diff are defined by a set of differential equations involving
the extended \W-generators.
\W-space is in some sense a bosonic version of the
superspace used in the covariant description of superconformal theories
\cite{Friedan}. However the dynamical fields are not present in the
defining relations of superconformal transformations in superspace.

The organization of the paper is as follows: In sect.\,2 we introduce
a set of covariant equations  used to construct the multi-time
{\W}$_N$ equations in \W-space. In sect.\,3 we consider the $N=3$ case
explicitly. Finite \W-Diff and the finite \W-transformations of the
extended generators are obtained in sect.\,4.
Conclusions and discussions are given in the last section.

%%%%%%%%%%%%%%%%%%%%%%%%%%%%%%%%%%%%%%%%%%%%%%%%%%%%%%%%%%%%%%%%%%%%%%

\section{ Covariant Formalism for Multi-time Equations }
\hspace{\parindent}%
% In the previous paper we have discussed how the multi-time equations
%are formulated using Riemannian geometry \cite{glasgow}.
Let us consider a system described by a set of fundamental
covariant equations \cite{glasgow},
\be
\label{cov}
\nabla_{\alpha} \nabla_{\beta} \, X^A(t^1,t^2,...,t^{N-1}) = 0,
\quad\quad\quad \alpha,\beta =t^1,\ldots,t^{N-1}.
\ee
We assume \bref{cov} to have $N$ independent solutions labeled by the
index $A=1,\ldots,N$.
Each $X^A$ is a scalar density of weight $h=1/N$ under general
coordinate transformations in $N-1$ dimensional space.
The covariant derivative is defined by
\be
\label{sccd}
\nabla_{\rho} X\,=\,
\partial_{\rho} X + h \Gamma^{\sigma}_{\sigma \rho} X\,=\,
(\partial_{\rho}  + h \Gamma_{ \rho})\,X,
\ee
and the second derivative is
\be
\label{second}
\nabla_{\nu} \nabla_{\rho}X \,=\, \partial_{\nu} \nabla_{\rho}X -
\Gamma^{\sigma}_{\rho \nu} \, \nabla_{\sigma}X + h
\Gamma^{\sigma}_{\sigma \nu} \, \nabla_{\rho}X \,=\,
\{\partial_{\nu}-{\bf A}_\nu \}_{\rho}^{\,\sigma} \, \nabla_\sigma X,
\ee
where
\be
\label{defA}
({\bf A}_\nu)_{\rho}^{\,\,\sigma}\,=\,
({\bf \Gamma}_\nu-h \Gamma_\nu)_{\rho}^{\,\,\sigma},
\,\,\,\,\,\,\,\,\,\,\,
({\bf \Gamma}_\alpha)_\beta^{\,\,\gamma}\,=\,
\Gamma_{\beta \alpha}^{\,\,\gamma},\,\,\,\,\,\,\,\,\,\,\,\,
{\Gamma}_\alpha\,=\,\Gamma_{\beta \alpha}^{\,\,\beta}\,=\,
{\rm tr}({\bf \Gamma}_\alpha).
\ee
The Christoffel symbol ${\bf \Gamma_\alpha}$ is regarded as $gl(N-1)$
valued gauge field and $\Gamma_\alpha$ as its trace.

 We can show that equation \bref{cov} is only consistent when the
Riemann-Christoffel curvature tensor vanishes.
Let us write the integrability condition of \bref{cov},
\be
\label{2-5}
\nabla_{[\mu} \nabla_{\nu]} \nabla_{\rho}X\,=\,
-({\bf F}_{\mu \nu})_\rho^{\,\lambda} \, \nabla_\lambda X \,-
\Gamma_{[\nu \mu]}^{\sigma} \nabla_\sigma \nabla_\rho X = 0.
\ee
Here  ${\bf F}$ is the field strength for ${\bf A}$,
\bea
\nonumber
 {\bf F}_{\alpha\beta}\, & \equiv &
\,\partial_\alpha {\bf A}_\beta-\partial_\beta {\bf A}_\alpha-
[{\bf A}_\alpha,\,{\bf A}_\beta]\,=\,
\\
\label{Rc}
& = & \,(\partial_\alpha {\bf\Gamma}_\beta-\partial_\beta {\bf
\Gamma}_\alpha-
[{\bf \Gamma}_\alpha,\,{\bf \Gamma}_\beta])\,-\,h\, {\rm tr}
(\partial_\alpha {\bf\Gamma}_\beta-\partial_\beta {\bf \Gamma}_\alpha-
[{\bf \Gamma}_\alpha,\,{\bf \Gamma}_\beta]).
\eea
The assumption of $N$ independent solutions for $X^A$
requires zero field strength, \,${{\bf F}=0}\,$.
By taking the trace of \bref{Rc}, the
vanishing of ${\bf F}$ implies zero Riemann-Christoffel
curvature,
\be
\label{zrc}
R_{\gamma \alpha \beta}^{\delta}\,=\,
\{\partial_\alpha {\bf\Gamma}_\beta-\partial_\beta {\bf \Gamma}_\alpha-
[{\bf \Gamma}_\alpha,\,{\bf \Gamma}_\beta]\}_\gamma^{\,\delta}\,=\,0.
\ee

Equation \bref{cov} also requires symmetric Christoffel
connections. In fact from \bref{second},
\be
\label{2-9}
0\,=\,[\nabla_\alpha,\nabla_\beta]X\,=\,
h(\partial_\alpha \Gamma_\beta-\partial_\beta \Gamma_\alpha)X\,-\,
\Gamma_{[\beta \alpha]}^{\,\gamma}\nabla_\gamma X.
\ee
The first term in the r.h.s. is just the trace of \bref{zrc} and
vanishes for the solutions. Thus the Christoffel symbols are
symmetric and there is no torsion,
\be
\Gamma_{\beta \alpha}^{\,\gamma}\,=\,
\Gamma_{\alpha \beta}^{\,\gamma}.
\ee

Since the Riemann-Christoffel tensor is zero there exists a set of flat
coordinates  $k^a$'s $\,(a=1,\ldots,N-1)$ such that the Christoffel
symbols vanish. Equation \bref{cov} has trivial solutions there. For
a generic system of coordinates we can write the Christoffel symbols as
\be
\label{christ}
    \Gamma_{\nu \mu}^{\rho}\,=\,( {\bf \Gamma}_\mu)_\nu^{\rho}\,=\,
(\partial_\mu
{\bf J})_\nu^{\,\,a}({\bf J}^{-1})_a^{\,\,\rho},\,\,\,\,\,\,\,\,\,\,\,
{\bf J}\,\in\,GL(N-1).
\ee
where
$
({\bf J})_\nu^{\,\,a}\,=\,\partial_\nu k^a(t^{\alpha}).
$\,\,
Due to the fact that $X$ is a scalar density of weight $h$   the
solution in generic coordinates  is
\be
\label{xsol}
X (t^{\alpha})\,=\,\ja^{-h}\left(C_a k^a
(t^{\alpha}) +C_N \right).
\ee
where $C_a$ and $C_N$ are integration constants.

The relation between the covariant system of equations \bref{cov}
and the multi-time equations for {\W}$_N$ \bref{multiDSeq} is
established through a suitable set of gauge conditions on the
Christoffel symbols.
The form of gauge-fixing conditions for general {\W}$_N$
have  been discussed in \cite{glasgow} and worked out for $N=3$
and 4 explicitly.
In the following sections we restrict ourselves to the {\W}$_3$
case and make the analysis of \W-Diff in detail.

%%%%%%%%%%%%%%%%%%%%%%%%%%%%%%%%%%%%%%%%%%%%%%%%%%
\section{Multi-time {\W}$_3$ equations}

\hspace{\parindent}%
We consider two-dimensional space with the local coordinates
$t \equiv t^1$ and $z \equiv t^2$. Let us impose the following
gauge-fixing conditions \cite{glasgow}:
\be
\label{gaugecond}
\gztt\,=\,1, \quad\quad\quad\quad \gttt=2 \gztz.
\ee
 Zero-curvature condition \bref{zrc} imposed on Riemann-Christoffel
curvature gives $4$ equations. It is used to express  $\gttz$ and $\gtzz$
in terms of independent Christoffel symbols, $\gztz$ and $\gzzz$,
\be
\label{ggttz}
{\gttz}\,=\,{\gztz}^2 + \gzzz - \pt\gztz,
\ee
and
\be
\label{ggtzz}
\gtzz\,=\,
%\gttz\,\gztz + \frac13\,\pt\gttz -  \frac23\,\pt\gzzz \,=\,
   {\gztz}^3 + \gztz\,\gzzz - \frac13\gztz\,\pt\gztz -
     \frac13\pt\gzzz - \frac13 \pt^2\gztz.
\ee
It also gives $z$-derivative of $\gztz$ and $\gzzz$
as functions of $\gztz$ and $\gzzz$ and their $t$ derivatives:
\bea
\label{ggztzz}
\pz\gztz\,& = & \,\frac13\,\pt(\gttz + \gzzz)\,=\,
   \frac13(2\,\gztz\,\pt\gztz + 2\,\pt\gzzz - \pt^2\gztz),
\\\nonumber
\pz\gzzz\,& = & \,\frac13\,(\,
   10\,{\gztz}^2\,\pt\gztz + 6\,\gzzz\,\pt\gztz - 2\,{(\pt\gztz)}^2
\\
& & \hspace{5mm} - 2\,\gztz\,\pt\gzzz +  2\,\gztz\,\pt^2\gztz +
        \pt^2\gzzz - 2\,\pt^3\gztz).
\label{ggzzzz}
\eea
In this gauge the covariant equations \bref{cov} are expressed as
\be
\label{xt3}
\pt^3 {\it X}(t,z)\,+\, T(t,z)\,\pt {\it X}(t,z)\, +\,
\left( W_3(t,z) + \frac12{\pt {T(t,z)}} \right) {\it X}(t,z)\,=\,0
\ee
and
\be
\label{xz}
\pz{\it X}(t,z)\,=\,\pt^2 {\it X}(t,z)\,+\,\frac23 \,T(t,z)\,{\it
X}(t,z).
\ee
They are the multi-time {\W}$_3$ equations
in which $T$ and $W_3$ are defined as functions of Christoffel symbols:
\bea
\label{TT}
& T(t,z)\,=\, {-2\,{\gztz}^2} - \gzzz +  2\,\pt\gztz \,=\,\gzzz-2\,
\gttz,
\\
\label{WW}
& {W_3(t,z)}\,=\,
   -{\gztz}^3 - \gztz\,\gzzz + \gztz\,\pt\gztz +
     \frac12\,\pt\gzzz\,=\,-\gtzz\,-\frac16\,\pt T(t,z).
\eea
The $z$-extension equations \bref{ggztzz} and \bref{ggzzzz}
for $\gztz$ and $\gzzz$ are translated to those for $T$ and $W_3$,
\bea
\label{Tz}
& \pz{T}(t,z)\,=\,{2\,\pt W_3}(t,z),
\\
\label{W3z}
& \pz{W_3}(t,z)\,=\,
   -\frac23\,T(t,z)\,\pt T(t,z) - \frac16 \pt^3 T(t,z).
\eea
They are nothing but the integrability conditions of the multi-time
{\W}$_3$ equations \bref{xt3} and \bref{xz}.

In sect.\,2 we have determined solutions of \bref{cov}
for $X^A$ and the Christoffel symbols in terms of $N-1$ arbitrary
functions $k^a$.
After imposing the gauge-fixing conditions \bref{gaugecond}
the functions $k^a$ are no longer arbitrary but must satisfy a set of
partial differential equations. The first gauge-fixing condition of
\bref{gaugecond}, $\,\gztt=1\,$, requires
\be
\label{jac}
\ja\,\equiv\,\pt k^1 \pz k^2 - \pt k^2 \pz k^1\,=\,
\pt k^1 \pt^2 k^2 - \pt k^2 \pt^2 k^1\,\equiv\,K_3.
\ee
Using the second gauge-fixing condition of \bref{gaugecond},
$\gttt=2\gztz$, we have
\be
\label{urk}
\pz k^a\,=\,\pt^2 k^a\,+\,Q\,\pt k^a,\quad\quad\quad
Q\,=\,-\frac23 {{\pt K_3}\over{K_3}},
\quad\quad\quad (a=1,2).
\ee

It is worth noticing that the extension equation for $k^a$
\bref{urk} shows
a global $SL(3)$ invariance. It is  non-linearly realized as
\be
\label{sl3}
k^a\,\,\longrightarrow\,\,{{k^b \,\tilde B_b^{\,a}+\tilde B_3^{\,a}}\over
{k^d \,\tilde B_d^{\,3}+\tilde B_3^{\,3}}},
\,\,\,\,\,\,\,\,\,\,\,a=1,2, \quad\quad
\tilde B\,\in\,SL(3)_{\rm global}. \ee
We also point out that the extension equation has a trivial set of
solution
\,$k^1=t,\,\, k^2=z+{{t^2}\over{2}}$\,  giving zero values for $T$ and
$W_3$.

The solutions of the multi-time {\W}$_3$
equations \bref{xt3}, \bref{xz},
\bref{Tz} and \bref{W3z} are now expressed in terms of $k^a$ satisfying
\bref{urk}.
Using \bref{xsol}, \bref{TT} and \bref{WW} we obtain:
\be
\label{solx}
X^A(t,z)\,=\,\{\,\, k^1 \, K_3^{-1/3},\,\,
 k^2 \, K_3^{-1/3},\,\,K_3^{-1/3}\,\,\},
\ee
\be
T(t,z)\,=\,
  {{\pt^2K_3}\over {K_3}}
-\frac43\,{{(\pt K_3)^2}\over {K_3^2}} +
\frac{1}{K_3}
    \left(\pt^2 k^1\,\pt^3 k^2  - \pt^2 k^2\,\pt^3 k^1 \right)
\ee
and
$$
W_3(t,z)\,=
- \frac16{{\pt^3K_3}\over {K_3}}
+ \frac56{{\pt K_3 \, \pt^2 K_3}\over{{{K_3}^2}}}
+ \frac56 \frac{\pt K_3}{K_3^2} \, \left(\pt^2 k^{1}\,\pt^3 k^{2}
-\pt^2 k^{2}\,\pt^3 k^{1} \right)
$$
\be
\label{solw3}
\hspace{-5mm}
- {{20}\over{27}}\,{{(\pt K_3)^3}\over {K_3^3}}
+ \frac{1}{2 \, K_3} \left(\pt^2 k^{2}\,\pt^4 k^{1} -
      \pt^2 k^{1}\,\pt^4 k^{2} \right).
\ee

These formulas coincide for $z=0$ with the
well-known expressions of the {\W}$_3$ generators as
obtained using, for example, the Wronskian method
\cite{marshakov}. If we use the extension equations for
$k^a$ \bref{urk} we
can obtain an alternative expression for $T(t,z)$ and $W(t,z)$ in terms
of $K_3$ in \bref{jac} only:
\bea
\nonumber
& T(t,z)\,=\,
-\frac 23\,{{(\pt K_3)^2}\over {K_3^2}} +
  \frac 12{{\pt^2K_3}\over {K_3}}
 -\frac 12 {{\pz K_3}\over {K_3}},
\\
\label{TWK}
& W_3(t,z)\,=\,
{{4}\over{27}}\,{{(\pt K_3)^3}\over {K_3^3}} -
  \frac14{{\pt K_3\,\pt^2K_3}\over
    {{{K_3}^2}}} +
  \frac{1}{12}{{\pt^3K_3}\over {K_3}} -
  {{5}\over{12}}\,{{\pt K_3{\pz K_3}} \over
    {{{K_3}^2}}} +
  {{1}\over{4}}\,{{\pt\pz K_3} \over
    {K_3}}.
\eea

The extension equation \bref{urk} is different from that discussed by
Gervais and Matsuo \cite{Matsuo}. They consider an equation
corresponding to \bref{urk} with $Q=0$. The condition that the
two-dimensional infinitesimal Diff preserve \bref{xt3} and
$\pz k^a=\pt^2 k^a$ cannot be written in a local way in terms of $T$
and $W_3$. It can be shown that \bref{urk} is the only possible form
of the extension equation satisfying this local property.

%%%%%%%%%%%%%%%%%%%%%%%%%%%%%%%%%%%%%%%%%%%%%%%%%%%%%

\section{\W-diffeomorphisms and Finite \W-Symmetry}

\hspace{\parindent}%
We have shown that the covariant equations \bref{cov} in the gauge
\bref{gaugecond} are the multi-time {\W}$_3$ equations
\bref{xt3}, \bref{xz}, \bref{Tz} and \bref{W3z} for $N=3$.
In this section we will show how the general coordinate transformations
that preserve the gauge conditions do generate the classical {\W}$_3$
transformations in the extended space.

Under general coordinate transformations $t=f({\tilde t},{\tilde
z}),\,z=g({\tilde t},{\tilde z})$ the
scalar density $X$ and the Christoffel symbols transform according to:
\be
\label{XF}
{\tilde X}({\tilde t},{\tilde z})=J^{-\frac13} X(t,z),\quad\quad\quad
J \equiv {{\p t}\over{\p {\tilde t}}}
   {{\p z}\over{\p {\tilde z}}}-{{\p t}\over{\p {\tilde z}}}{{\p z}\over
{\p  {\tilde t}}},
\ee
\be
\label{GF}
\tilde \Gamma^{\alpha}_{\beta\gamma}({\tilde t},{\tilde z})=
\left({{\p t^{\mu}}\over{\p {\tilde t}^{\beta}}}
 {{\p t^{\nu}}\over{\p {\tilde t}^{\gamma}}} \Gamma^{\rho}_{\mu\nu}(t,z)+
{{\p^2 t^{\rho}}\over{\p {\tilde t}^{\beta}\p {\tilde
t}^{\gamma}}}\right) {{\p {\tilde t}^{\alpha}}\over{\p t^{\rho}}}.
\ee
{}From the requirement that general coordinate transformations keep
the gauge conditions \bref{gaugecond} invariant we find a set of
equations to be satisfied by the transformation functions $f({\tilde
t},{\tilde z})$ and $g({\tilde t},{\tilde z})$:
\bea
\nonumber
& f'=\ddot f - \frac23 \dot f {{\dot J}\over{J}} - \frac23 \dot f \dot g
T(t,z) -\dot g^2 V_3(t,z),
\\
\label{fext}
& g'=\ddot g - \frac23 \dot g {{\dot J}\over{J}} + \dot f^2
 +\frac13 \dot g^2 T(t,z).
\eea
Here $\dot f={{\p f({\tilde t},{\tilde z})}\over{\p {\tilde t}}}$,
$f'={{\p f({\tilde t},{\tilde z})}\over{\p {\tilde z}}}$ and so on.
Equations \bref{TT} and \bref{WW} have been used to give the
expressions of $T$ and $V_3 \equiv W_3 + \frac16 \pt T$ in terms of
Christoffel symbols.
Then we define a \W-Diff as a general coordinate transformation
$t=f({\tilde t},{\tilde z}),\,z=g({\tilde t},{\tilde z})$ satisfying
equations \bref{fext} for given $T$ and $V_3$.
In contrast to the conformal and superconformal transformations they
depend on the connections $T$ and $V_3$. In other words, the
two-dimensional coordinate \W-transformations cannot be performed
independently of the \W-generators of the system.

The finite transformations of $X$ are given by \bref{XF} with the
Jacobian determined from \bref{fext}:
\be
J({\tilde t},{\tilde z})=\dot f \ddot g -\ddot f \dot g +\dot f^3 +
\dot f \dot g^2 T(t,z) + \dot g ^3 V_3(t,z).
\ee
The finite transformations of the extended \W-generators are obtained
from those of Christoffel connections,
$$
\tilde T ({\tilde t},{\tilde z})={{1}\over{J}}\left(\dot f g''-f''
\dot g -2 (\dot
f' g'-f'\dot g') +3 \dot f f'^2 +(\dot f g'^2+2 f'\dot g g')T(t,z) + 3
\dot g g'^2 V_3(t,z)\right),
$$
\be
\label{V3tilde}
\tilde V_3 ({\tilde t},{\tilde
z})={{1}\over{J}}\left(f'g''-f''g'+f'^3+f'g'^2 T(t,z)+g'^3 V_3
(t,z) \right).
\ee
Notice that $f$ and $g$ are not arbitrary functions but
satisfy \bref{fext} and depend implicitly
on $T$ and $V_3$. Thus the $\tilde T$ and $\tilde V_3$ in \bref{V3tilde}
have non-linear dependence on $T$ and $V_3$.

%%%%%%%%%%%GLOBAL%%%%%%%%%%%%%%%%%%%%%%%%%%%%%%%%%%%%%%%%%%%

\vspace{5mm}

Let us consider a coordinate system on which $T=0$ and $W_3=0$ and
consider a subset of \W-Diff which can be performed
on this coordinate system. It can be constructed using solutions
$k^a$ of the extension equations \bref{urk} by
\be
\label{kfg}
f=k^1 \quad\quad\quad {\rm and} \quad\quad\quad g=k^2 - \frac12(k^1)^2.
\ee
The Jacobian $J$ of this  particular transformation is $K_3$ given in
\bref{jac} and the transformed generators $T$ and $W_3$ take the same form
as \bref{TWK}. They can be rewritten in a more compact form
as
\be
T=\frac32 J^{1/3} (\pz - \pt^2)J^{-1/3},\quad\quad\quad
W_3=\frac12 J^{1/3} (\pt^3-3 \pt\pz)J^{-1/3}+\frac12 \pt T.
\label{TWJ}
\ee

Finite global \W-Diff are defined as those leaving
the values $T=0$ and $W_3=0$ invariant. They have the general
form
\be
\label{glsl3}
t\,={{a{\tilde t}+b({\tilde z}+{{{\tilde t}^2}\over 2})+c} \over{q{\tilde
t}+r({\tilde z}+{{{\tilde t}^2}\over 2})+s}},\quad\quad\quad z+{{t^2}\over
2}\,={{m{\tilde t}+n({\tilde z}+{{{\tilde t}^2}\over 2})+p}
\over{q{\tilde t}+r({\tilde z}+{{{\tilde t}^2}\over 2})+s}},
\ee
$$
\pmatrix{a&b&c\cr m&n&p\cr q&r&s\cr} = {\rm constant} \in
\,SL(3).
$$
The Jacobian of the transformations \bref{glsl3} gives
vanishing $T$ and $W_3$ in \bref{TWJ}. This property
suggests to consider the expressions \bref{TWJ} as the associated
\W-Schwarzians.

Note that \bref{sl3} is a projective realization of an $SL(3)$
transformation law which is generalizing the projective $SL(2)$ (M{\"o}bius)
transformation in the standard conformal theories. Therefore it is
rather natural to consider the two-dimensional space where we describe
{\W}$_3$ as being $RP^2$.

 %%%%%%%%%%%%Infinitesimal%%%%%%%%%%%%%%%%%%%%%%%%%%%%%%%%%

\vspace{5mm}

For the infinitesimal general coordinate transformations
$\delta t=\et(t,z)$ and  $\delta z=\ez(t,z)$ the conditions of \W-Diff
\bref{fext} are expressed as:
\bea
\label{epszz}
&&\pz{\it \epsilon^z}(t,z)\,=\,
   2\,\pt{\it \epsilon^t}(t,z) + \pt^2{\it \epsilon^z}(t,z),
\\
\label{epstz}
&&\pz{\it \epsilon^t}(t,z)\,=\,- \frac23 T(t,z)\,\pt{\it
\epsilon^z}(t,z) - \frac23\pt\pz{\it \epsilon^z}(t,z) + \frac13
     \pt^2{\it \epsilon^t}(t,z).
\eea
The infinitesimal \W-Diff transformations of the extended
\W$_3$-generators are
\be
\label{delT}
\delta T\,=\,
 {\it \epsilon^z}\,\pz T  +  {\it \epsilon^t}\,\pt T +
  T\,\pz{\it \epsilon^z} + \pz^2 {\it \epsilon^z} +
3\,V_3 \,\pt{\it \epsilon^z} - 2\,\pt\pz{\it
\epsilon^t} \ee
and
\be
\label{delWhat}
\delta {V_3}\,=\,
 {\it \epsilon^z}\,\pz{V_3} + {\it \epsilon^t}\,\pt{V_3}+
  T\,\pz{\it \epsilon^t}+ 2\,{V_3}\,\pz{\it \epsilon^z} -
  \pz^2{\it \epsilon^t} - {V_3}\,\pt{\it \epsilon^t}.
\ee

To see the relations of these transformations with the standard {\W}$_3$
transformations in one dimension we express all $z$-derivatives in terms of
$t$-derivatives using \bref{Tz}, \bref{W3z}, \bref{epszz} and
\bref{epstz}. The resulting transformations are:
\bea
\label{delTalp}
& \delta T(t,z)\,=\, {\it \alpha}\,\pt T +
2\,\pt{\it \alpha}\,T\, +\, 2\,\pt^3 {\it \alpha}+
\,2\,{\it \rho}\,\pt W_3 +
  3\,(\pt\,{\it \rho})\,W_3,
\\\nonumber
& \delta W_3(t,z)\,=\,{\it \alpha}\,\pt W_3\,+\,
3\,\pt{\it \alpha}W_3\,
-{\it \rho}\,\left( {{2\,T\,\pt T}\over3} + {{\pt^3 T}\over 6} \right)
- \pt {\it \rho}\,\left({{2\,{{T}^2}}\over 3}+{{3\,\pt^2 T}\over 4}
\right)
\\
& -\frac54 \,(\pt^2{\it \rho}) \,\pt T\,-
\frac56\, (\pt^3 {\it \rho})\,T  -   \frac16{\pt^5{\it \rho}},
\\\nonumber
& \delta X(t,z)\,=\, {\it \alpha}\,\pt X\,
-\,\pt{\it \alpha}\,X
+ \,{\it \rho}\,(\,\pt^2 X\,+\,\frac23{\,T\,X}) -
\\
\label{delXrho}
& \frac12 {{(\pt{\it \rho})\,\,\pt X}} + \frac16{{\,(\pt^2{\it \rho})}\,\,X}.
\eea
where $\alpha(t,z) \equiv \et (t,z) + \frac12 \pt \ez (t,z)$ and $\rho
(t,z) \equiv \ez(t,z)$.

The transformations \bref{delTalp}-\bref{delXrho} are reduced to the
classical infinitesimal {\W}$_3$ transformations by putting
$z=0$ and considering $\alpha(t,0)$ and $\rho(t,0)$ as
arbitrary transformation functions of $t$.
The parameter $\alpha$ generates $t$
diffeomorphism transformations
under which $T$, $W_3$ and $X$ transform respectively
as weight 2 quasi-primary, weight 3 and weight $-1$ primary fields.
The transformations generated by $\rho$ are the well-known
{\W}$_3$ transformations.

The algebra of two infinitesimal \W-Diff is given by
\be
 [\delta_{\epsilon_1},\,\delta_{\epsilon_2}]\,=\,\delta_{\epsilon_3},
\quad\quad\quad \epsilon_3^\mu=\epsilon^\nu_2 \p_\nu \epsilon^\mu_1
                          -\epsilon^\nu_1 \p_\nu \epsilon^\mu_2
+ \delta_1 \epsilon^\mu_2- \delta_2 \epsilon^\mu_1.
\ee
Here the last two terms of $\epsilon_3$ are contributions coming
from the $T$ dependence of $\epsilon$'s through \bref{epszz} and
\bref{epstz}. They satisfy, for example:
\be
\pz{(\delta \epsilon^t)}\,=\,- \frac23 (T\,\pt{(\delta \epsilon^z)}
+(\delta T) \pt \epsilon^z) - \frac23\pt\pz{(\delta \epsilon^z)} +
\frac13
     \pt^2{(\delta \epsilon^t)}.
\ee
Taking this into account it can be shown
that the transformation with parameter $\epsilon_3$ is also
a \W-Diff because it satisfies equations \bref{epszz} and \bref{epstz}.
Therefore we can say that the \W-Diff have a composition law, at
least locally, forming a quasi-group.

The existence of a composition law enables us to define the \W-surface
in general. Let us consider the manifold $M=RP^2$ with a flat affine
connection $\Gamma$. We define a \W-neighborhood as an ordinary one
supplemented with the conditions $\gztt=1$ and $\gttt-2 \gztz=0$. These
conditions single out some special parametrizations of each patch of
$M$. General two-dimensional Diff will not preserve these conditions
but \W-Diff, as
defined above, do preserve them. Hence we can define the \W-surface as a
collection of \W-neighborhoods patched together by \W-Diff.

%%%%%%%%%Conclusions%%%%%%%%%%%%%%%%%%%%%%%%%%%%%%%%%%%%%%%

\section{Conclusion}

\hspace{\parindent}%
We have studied {\W}$_3$ transformations in the language of Riemannian
geometry. The \W-generators are given as particular combinations of the
Christoffel symbols in a suitable gauge. We have seen that \W-Diff
preserving the gauge-fixing conditions depend in general on $T$
and $W_3$. In general the finite transformations of $T$ and $W_3$ are
expressed as non-linear functions of the fields. Explicit expressions
for the finite global transformations and the \W-Schwarzians have been
given. We have also indicated how to recover the well-known
one-dimensional {\W}$_3$-transformations from our results.

It is desirable that the gauge-fixing condition \bref{gaugecond} is
interpreted geometrically in order to have a better understanding of
\W-surfaces. We would like to emphasize that the definition of \W-Diff
\bref{fext} given above may be valid for more general cases and we
believe that it will be useful, for example, in the study of the
classical limit of the {\W}$_3$ string theory.

\vspace{10mm}

{\bf Acknowledgements:}
K.K. would like to thank Dept.\,of ECM (Barcelona) for their hospitality
during his stay.
J.R. thanks the Spanish Ministry of Education and the British Council
for financial support.
J.H. acknowledges a fellowship from Generalitat de Catalunya.
This work has been partially supported by CYCYT under  contract number
AEN93-0695 and by Commission of the European Communities
contract CHRX-CT93-0362(04).

%%%%%%%%%%%%%%%%%%%%%%%%%%%%%%%%%%%%%%%%%%%%%%%%%%%%%%%%%%%%%%%%

\end{document}